\documentclass[aps,prd,showpacs,showkeys,preprint,floatfix,eqsecnum,superscriptaddress,notitlepage]{revtex4-1}

\usepackage{amssymb}
\usepackage{amsmath}
\usepackage{amsbsy}
\usepackage{bm}
\usepackage{color}
\usepackage{graphicx}
\usepackage{slashed}

\newcommand{\bq}{\begin{eqnarray}}
\newcommand{\eq}{\end{eqnarray}}
\newcommand{\bqn}{\begin{eqnarray*}}
\newcommand{\eqn}{\end{eqnarray*}}
\newcommand{\RR}{{\bf R}}
\newcommand{\KK}{{\bf K}}
\newcommand{\kk}{{\bf k}}
\newcommand{\rr}{{\bf r}}

\newcommand{\pp}{{\bf p}}
\newcommand{\cc}{{\cal C}}

\begin{document}
\title{White-dwarfs equation of state and structure: The effect of
  temperature}  

\author{Riccardo Fantoni}
\email{rfantoni@ts.infn.it}
\affiliation{Universit\`a di Trieste, Dipartimento di Fisica, strada
  Costiera 11, 34151 Grignano (Trieste), Italy}

\date{\today}

\begin{abstract}
We study the effect of having a finite temperature on the equation of
state and structure of a white dwarf. In order to keep the treatment
as general as possible we carry on our discussion for ideal quantum
gases obeying to both the Fermi-Dirac and the Bose-Einstein
statistics even if we will only use the results for the free electron
gas inside a white dwarf. We discuss the effect of temperature on the
stability of the star and on the Fermi hole. 
\end{abstract}



\maketitle
\section{Introduction}
\label{sec:introduction}

A {\sl white dwarf} below the regime of neutron drip, at mass densities
less than $4\times 10^{11}\text{g~cm}^{-3}$, are stars which emit light of
a white color due to their relatively high surface temperature of
about $10^4 \text{K}$. Because of their small radii $R$, luminous white
dwarfs, radiating away their residual thermal energy, are
characterized by much higher effective temperatures, $T$, than normal
stars even though they have lower luminosities (which varies as
$R^2T^4$). In other words, white dwarfs are much ``whiter'' than
normal stars, hence their name \cite{Balian99,Silbar04,Jackson05}.

White dwarfs life begins when a star dies, they are therefore {\sl
  compact objects} \cite{Shapiro-Teukolsky}. Star death begins when
most of the nuclear fuel has been consumed. White dwarfs has about one
solar mass $M_\odot$ with characteristic radii of about $5000
\text{km}$ and mean densities of around $10^6 \text{g~cm}^{-3}$. They
are no longer burning nuclear fuel and are slowly cooling down as they
radiate away their residual thermal energy. 

They support themselves against gravity by the pressure of cold
electrons, near their degenerate, zero temperature, state. 
In 1932 L. D. Landau \cite{Landau1932} presented an elementary
explanation of the equilibrium of a white dwarf which had been
previously discovered by Chandrasekhar in 1931 
\cite{Chandrasekhar1931,Chandrasekhar1931a,Chandrasekhar1931b}
building on the formulation of the Fermi-Dirac statistics in August
1926 \cite{Dirac1926} and the work of R. H. Fowler in December 1926
\cite{Fowler1926}, on the role of the {\sl electron degeneracy
  pressure} to keep the white dwarf from gravitational
collapse. Landau explanation can be found in \S 3.4 of the book of
Shapiro and Teukolsky \cite{Shapiro-Teukolsky}, and fixes the
equilibrium maximum mass of the white dwarf to $M_\text{max}\sim
1.5M_\odot$. Whereas Chandrasekhar result was
$M_\text{Ch}=1.456M_\odot$ for completely ionized matter made of
elements with a ratio between mass number and atomic number equal to
2. Strictly speaking one would have a matter made of a fluid of
electrons and a fluids of nuclei. In the work of Chandrasekhar the
fluid of electrons is treated as an ideal gas where the electrons are
not interacting among themselves and the nuclei thousands times
heavier are neglected. 

Despite the high surface temperature these stars are still considered
cold, however, because on a first approximation temperature does not
affect the equation of state of its matter. White dwarfs are described
as faint stars below the main sequence in the Hertzsprung-Russell
diagram. In other words, white dwarfs are less luminous than
main-sequence stars of corresponding colors. While slowly cooling, the
white dwarfs are changing in color from white to red and finally to
black. White dwarfs can be considered as one possibility of a final
stage of stellar evolution since they are considered static over the
lifetime of the Universe. 

White dwarfs were established in the early $20^\text{th}$ century and
have been studied and observed ever since. They comprise an estimated
$3\%$ of all the stars of our galaxy. Because of their low luminosity,
white dwarfs (except the very nearest ones) have been very difficult
to detect at any reasonable distance and that is why there was very
little observational data supporting the theory in the time of them
being discovered. The companion of Sirius, discovered in 1915 by
W. S. Adams \cite{Adams1915,Adams1925}, was
among the earliest to become known. The cooling of white dwarfs is not
only a fascinating phenomenon but in addition offers information of
many body physics in a new setting since the circumstances of an
original star can not be built up in a laboratory. More over, the
evolution and the equation of state for white dwarfs can be useful on
Earth providing us more understanding of matter and physics describing
the Universe. 

In this work, we discuss how the Chandrasekhar analysis at zero
temperature should be changed in order to take into account the effect
of having a quantum ideal gas at finite (non-zero) temperature. For
the sake of generality we will treat in parallel the case of the Fermi
and the Bose ideal gases. Even if only the Fermi case is appropriate
for the description of the white dwarf interior made of ionized matter
characterized by a sea of free cold electrons (as Chandrasekhar did,
we will neglect the Coulomb interaction between the electrons and
disregards the nuclei in order to keep the treatment analytically
solvable. We will also use Newtonian gravity to study the star
stability disregarding general relativistic effects). At the typical
surface temperature and density of a white dwarf the momentum thermal
average fraction of particles having momentum $\hbar\kk$ and a full
relativistic dispersion relation ($\cc_k/\cc_0$ where $\cc_k$ is given
by Eq. (\ref{ck}) below) varies appreciably over a $k$ range which is
a fraction of $0.933$ \footnote{This value 
will get smaller as the star cools down in view of Eq. (\ref{denerf})
and eventually become close to zero as the momentum thermal average
fraction approaches a step function}
of the $k$ range where it is different from zero. So we generally
expect the effect of temperature to play a role on the behavior of the
ideal quantum gas. We will pursue our analysis for both the
thermodynamic properties, as the validity of the various polytropic
adiabatic equation of state as a function of density, and for the
structural properties, as the Fermi hole.     

The paper is organized as follows: In section
\ref{sec:iqg-thermodynamics} we review the thermodynamic properties
of the ideal quantum gases at finite temperatures. This section
contains three subsections, in the first one \ref{sec:rehd} we discuss
the importance of a full relativistic treatment at high densities, in
the second one \ref{sec:ood} we discuss the onset of quantum
statistics as the star collapses, and in the third one \ref{sec:Cl} we
present the revised Chandrasekhar analysis. In the second section
\ref{sec:iqg-structure} we present our study of the structure of the
ideal quantum gases at finite temperature and in the full relativistic
regime. 

\section{The thermodynamics of the ideal quantum gas}
\label{sec:iqg-thermodynamics}

We want to find the thermodynamic grand potential of a system of many
free fermions or bosons with a rest mass $m$ in thermodynamic
equilibrium at an inverse temperature $\beta=1/k_B T$. 
 
The Hamiltonian of the system is 
\bq
{\cal H}=\sum_i(-\hbar^2c^2\Delta_i+m^2c^4)^{1/2}~,
\eq
with $\Delta$ the Laplacian and $c$ the speed of light.

Assuming the many particles are distinguishable (Boltzmannons) the
density matrix operator, $\hat{\rho}_D$, satisfies to the Bloch
equation 
\bq \label{rhoDf}
\frac{\partial \hat{\rho}_D(\beta)}{\partial \beta}&=&
-{\cal H}\hat{\rho}_D(\beta)~,\\
\hat{\rho}_D(0)&=&{\cal I}~,
\eq
where ${\cal I}$ is the identity operator. The solution of
Eq. (\ref{rhoDf}) in coordinate representation
$\RR=(\rr_1,\ldots,\rr_N)$, where $\rr_i$ is the position of $i$th
spinless particle in the three dimensional space, has the following
solution  
\bq \label{prop}
\rho_D(R_0,R_1;\beta)=\langle R_0\mid e^{-\beta{\cal H}}\mid R_1\rangle=
\int \frac{d\KK}{(2\pi)^{3N}}e^{-i\KK\cdot(\RR_0-\RR_1)}
e^{-\beta\sum_i(\hbar^2c^2\kk_i^2+m^2c^4)^{1/2}}~,
\eq
where $\KK=(\kk_1,\ldots,\kk_N)$ and $\RR_n=(\rr_1^n,\ldots,\rr_N^n)$. 
A very simple calculation yields the propagator $\rho_D$ in closed
form. The result can be cast in the following form
\bq
\rho_D&=&\prod_i{\cal R}({\rr_i}^1,{\rr_i}^0)~,
\eq
where ${\cal R}$ in one dimension is
\bq
{\cal R}_\text{1d}(\rr^1,\rr^0)&=&\frac{mc^2\beta}{\pi\Psi^{1/2}}
K_1\left(\frac{mc}{\hbar}\Psi^{1/2}\right)~,
\eq 
where $\Psi=(\rr^1-\rr^0)^2+(\hbar c\beta)^2$ and $K_\nu$ is
the familiar modified Bessel functions of order $\nu$. In three
dimensions we thus find  
\bq 
{\cal R}(\rr^1,\rr^0)&=&-\frac{1}{2\pi|\rr^1-\rr^0|}
\frac{d{\cal R}_\text{1d}(\rr^1,\rr^0)}{d|\rr^1-\rr^0|}\\ \nonumber
&=&\frac{mc^2\beta}{4\pi^2\Psi^{3/2}}
\left[\frac{mc}{\hbar}\Psi^{1/2}K_0\left(\frac{mc}{\hbar}\Psi^{1/2}\right)+
2K_1\left(\frac{mc}{\hbar}\Psi^{1/2}\right)+
\frac{mc}{\hbar}\Psi^{1/2}K_2\left(\frac{mc}{\hbar}\Psi^{1/2}\right)\right]~,
\eq
Note that for the non relativistic gas, when ${\cal
  H}=-\lambda\sum_i\Delta_i$, $\rho_D$ would have been the usual
Gaussian $\Lambda^{-3N}e^{-(\RR_1-\RR_0)^2/4\lambda\beta}$, with
$\lambda=\hbar^2/2m$ and $\Lambda=\sqrt{4\pi\beta\lambda}$ the de
Broglie thermal wavelength.  

Taking care of the indistinguishability of the particles we can
describe a system of bosons and fermions with spin $s=(g-1)/2$ through
density matrices, $\hat{\rho}_{B,F}$, that are obtained from the
distinguishable one opportunely symmetrized or antisymmetrized,
respectively. The corresponding grand canonical partition functions
can then be found through a standard procedure \cite{FeynmanFIP} from
$\Theta_{B,F}=e^{-\beta\Omega_{B,F}}=\sum_{N=0}^\infty
Z^N_{B,F}e^{N\mu\beta}$ where $Z^N_{B,F}=e^{-\beta F_{B,F}^N}$ is the
trace of $\hat{\rho}_{B,F}$. Here $\mu=(\ln z)/\beta$ is the chemical
potential, $F$ is the Helmholtz free energy, and $\Omega$ is the grand
thermodynamic potential.

If $V$ is the volume occupied by the system of particles, the
pressure is given by $P=-\Omega/V$, and the average number of
particles, $N=n V=-z\partial\beta\Omega/\partial z$, where $n$ is the
number density. We find for bosons 
\bq \label{Prb}
\beta P&=&\frac{gm^2c}{2\pi^2\beta\hbar^3}\sum_{\nu=1}^\infty
\frac{z^\nu}{\nu^2}K_2(\beta mc^2\nu)~,\\ \label{denrb}
n&=&\frac{gm^2c}{2\pi^2\beta\hbar^3}\sum_{\nu=1}^\infty
\frac{z^\nu}{\nu}K_2(\beta mc^2\nu)~,
\eq
and for fermions
\bq \label{Prf}
\beta P&=&\frac{gm^2c}{2\pi^2\beta\hbar^3}\sum_{\nu=1}^\infty
\frac{(-1)^{\nu-1}z^\nu}{\nu^2}K_2(\beta mc^2\nu)~,\\ \label{denrf}
n&=&\frac{gm^2c}{2\pi^2\beta\hbar^3}\sum_{\nu=1}^\infty
\frac{(-1)^{\nu-1}z^\nu}{\nu}K_2(\beta mc^2\nu)~.
\eq
Clearly in the zero temperature limit ($\beta\to\infty$) these reduce
to (see \S 2.3 of Ref. \cite{Shapiro-Teukolsky} and our appendix
\ref{app:1})  
\bq \label{Prdf1}
P&=&\frac{g}{2}\frac{mc^2}{\slashed{\lambda}^3}\phi(x)~,\\ \label{Prdf2}
n&=&\frac{g}{2}\frac{x^3}{3\pi^2\slashed{\lambda}^3}~,\\ \label{Prdf3}
\phi(x)&=&\frac{1}{8\pi^2}\left[x\sqrt{1+x^2}\left(
\frac{2}{3}x^2-1\right)+\ln\left(x+\sqrt{1+x^2}\right)\right]~,
\eq
where $\slashed{\lambda}=\hbar/mc$, with $m$ the electron mass, is the
electron Compton wavelength. 

We can then introduce the polylogarithm, $b_\mu$, of order $\mu$ and
the companion $f_\mu$ function,
\bq
b_{\mu}(z)&=&\sum_{\nu=1}^\infty
\frac{z^\nu}{\nu^{\mu}}~,\\
f_{\mu}(z)&=&\sum_{\nu=1}^\infty
\frac{(-1)^{\nu-1}z^\nu}{\nu^{\mu}}=-b_{\mu}(-z)=
\left(1-2^{1-x}\right)b_\mu(z)~.
\eq

At finite temperatures, in the extreme relativistic case, we find for
bosons   
\bq
\beta P&=&\frac{g}{\pi^2(\beta\hbar c)^3}b_{4}(z)~,\\
n&=&\frac{g}{\pi^2(\beta\hbar c)^3}b_{3}(z)~,
\eq
where we used the property $zdb_\mu(z)/dz=b_{\mu-1}(z)$, and for
fermions 
\bq 
\beta P&=&\frac{g}{\pi^2(\beta\hbar c)^3}f_{4}(z)~,\\ \label{denerf}
n&=&\frac{g}{\pi^2(\beta\hbar c)^3}f_{3}(z)~,
\eq
In agreement with \S 61 of Landau \cite{LandauCTP5}. 
And in the non relativistic case, we find for bosons
\bq
\beta P&=&\frac{g}{\Lambda^3}b_{5/2}(z)~,\\ \label{dennrb}
n&=&\frac{g}{\Lambda^3}b_{3/2}(z)~,
\eq
and for fermions
\bq
\beta P&=&\frac{g}{\Lambda^3}f_{5/2}(z)~,\\
n&=&\frac{g}{\Lambda^3}f_{3/2}(z)~,
\eq
In agreement with  \S 56 of Landau \cite{LandauCTP5}.
Recalling that the internal energy of the system is given by
$E=-\partial\ln\Theta/\partial\beta$ we find in the extreme
relativistic case $E=3PV$ and in the non relativistic case $E=3PV/2$. 
At very low density $n$, and high temperature $T$, when
$n/T^{3/2}$ is very small, $b_{3/2}(z)\approx f_{3/2}(z)$ is very
small and $z$ is also very small. In this case 
$b_{3/2}(z)\approx b_{5/2}(z)\approx f_{3/2}(z)\approx
f_{5/2}(z)\approx z$ and we find for the quantum gas $E/V\approx
(3/2)K_B Tn$. That is the non relativistic classical limit.
For the bosons, as the temperature gets small at fixed density
$b_{3/2}(z)$ increases (see Eq. (\ref{dennrb})) and $z$ gets close
to $1$. $b_\mu(z)$ is a monotonically increasing function of $z$ which
is only defined in $0\le z\le 1$, so the bosons ideal gas must
have a chemical potential less than
zero. $b_{3/2}(1)=\zeta(3/2)\approx 2.612$ and  
$b_{5/2}(1)=\zeta(5/2)\approx 1.341$ where $\zeta$ is the Riemann zeta 
function. The temperature
$T_c=\frac{2\pi\hbar^2}{mk_B}\left(\frac{n/g}{\zeta(3/2)}\right)^{2/3}$  
at which $z=1$ is called the {\sl critical temperature} for the
Bose-Einstein condensation in the non relativistic case. For $T<T_c$
the number of bosons with 
energy greater than zero will then be $N_>=N(T/T_c)^{3/2}$. The rest
$N_0=N[1-(T/T_c)^{3/2}]$ bosons are in the lowest energy state, {\sl
  i.e.} have zero energy. For the fermions the activity is allowed to
vary in $0\le z<\infty$ and the functions $f_\mu(z)$ can be extended
at $z>1$ by using the following integral representation
$f_x(z)=[\int_0^\infty dy\,y^{x-1}/(e^y/z+1)]/\Gamma(x)$, where
$\Gamma$ is the usual gamma function.   

Given the entropy $S=-\partial\Omega/\partial T$ we immediately see
that, in both the extreme relativistic and the non relativistic cases,
$S/N$ must be a homogeneous function of order zero in $z$ and 
that along an adiabatic process ($S/N$ constant) we must have $z$
constant. Then on an adiabatic, in the extreme relativistic case,
$P\propto n^{1+1/3}$, a polytrope of index $3$, and in the non
relativistic case, $P\propto n^{1+2/3}$, a polytrope of index
$3/2$. This conclusions clearly continue to hold at zero temperature
when $z\to\infty$ and the entropy is zero.

\subsection{Relativistic effects at high density in a gas of fermions}
\label{sec:rehd}

The thermal average fraction of particles having momentum
$\pp=\hbar\kk$ is given by
\bq \label{ck}
\cc_k=\frac{g}{N}\frac{1}{e^{\beta[\epsilon(k)-\mu]}-\xi}=
\frac{g}{N\xi}b_0\left(\xi z e^{-\beta\epsilon_k}\right)~,~~~
V\int\frac{d\kk}{(2\pi)^3}\cc_k=1~,
\eq
where $\xi=+1,-1,0$ refer to the Bose, Fermi, and Boltzmann gas
respectively. 

In a degenerate ($T=0$) Fermi gas we can define a Fermi energy as
$\epsilon_F=\mu=\sqrt{p_F^2c^2+m^2c^4}$, in terms of the Fermi momentum
$p_F$. From Eq. (\ref{ck}) follows that the thermal average
fraction of particles having momentum $\pp=\hbar\kk$ is
$\cc_k=(g/N)\Theta[\mu-\epsilon(k)]$, where $\Theta$ is the Heaviside
unit step function and $\epsilon(k)=\sqrt{\hbar^2k^2c^2+m^2c^4}$ is
the full relativistic dispersion relation. We will then have for the
density
\bq
n=\frac{g}{h^3}\int_0^{p_F}4\pi p^2\,dp=\frac{4\pi g}{3h^3}p_F^3~.
\eq
We then see immediately that at high density the Fermi momentum is
also large and as a consequence the Fermi gas becomes relativistic. On
the contrary the degenerate Bose gas will undergo the Bose Einstein
condensation and have all the particles in the zero energy state.

At finite temperature from the results of the previous section we find
that since $f_\mu(z)$ is a monotonously increasing function of $z$
then at large density $n$ also $z$ is large and at fixed temperature
this implies that the chemical potential $\mu$ is also large. In view
of Eq. (\ref{ck}) this means that in the gas there are fermions of
ever increasing momentum so that a relativistic treatment becomes
necessary.

\begin{figure}[htbp]
\begin{center}
\includegraphics[width=10cm]{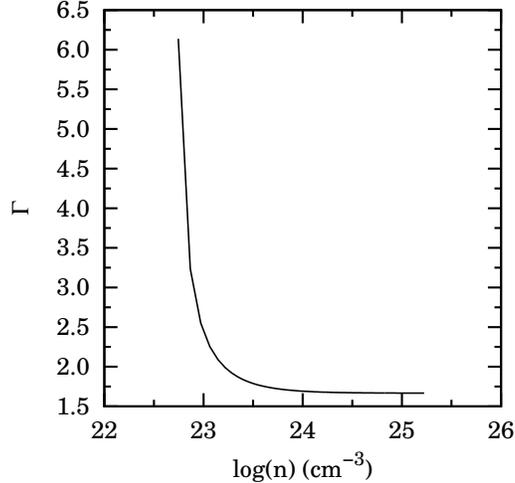}
\end{center}  
\caption{The exponent $\Gamma=d\ln P/d\ln n$ for the adiabatic
  full relativistic equation of state as a function of
  density. We chose a temperature $T=20000\text{K}$ and zero entropy,
  $g=2$, and $m$ is the mass of an electron. $n$ is in
  $\text{cm}^{-3}$.} 
\label{fig:eos}
\end{figure}

From Eqs. (\ref{Prf}) and (\ref{denrf}) it is possible (see appendix
\ref{app:1}) to extract the 
full relativistic adiabatic equation of state as a function of
temperature and observe the transition from the low density regime to
the high density extreme relativistic one. In Fig. \ref{fig:eos} we
show the exponent $\Gamma=d\ln P/d\ln n$ for the adiabatic full
relativistic equation of state as a function of density. For the sake
of the calculation it may be convenient to use natural units
$\hbar=c=k_B=1$. From the figure we see how at high density (which
implies high activity) $\Gamma\to 4/3$. This figure should be compared
with Fig. 2.3 of Ref. \cite{Shapiro-Teukolsky} for the degenerate
Fermi gas. In particular we see how at a temperature of
$T=20000\text{K}$ the Fermi gas can be considered extremely
relativistic already at an electron number density $n\gtrsim
10^{25}\text{cm}^{-3}$. While we know (see
Ref. \cite{Shapiro-Teukolsky} and Eqs. (\ref{Prdf1})-(\ref{Prdf3}))
that the completely degenerate gas becomes extremely relativistic for
$n\gtrsim 10^{31}\text{cm}^{-3}$. 
%

\subsection{The onset of quantum statistics}
\label{sec:ood}

For a spherically symmetric distribution of matter, the mass interior
to a radius $r$ is given by
\bq \label{masss}
m(r)=\int_0^r\rho4\pi {r'}^2\,dr'~,~~~\mbox{or}~~~
\frac{dm(r)}{dr}=4\pi r^2\rho~.
\eq
Here, since as we are considering non relativistic matter made of
completely ionized elements of atomic number $Z$ and mass number $A$,
$\rho=\rho_0=\mu_em_un$ is the rest mass density with $\mu_e=A/Z$ 
the mean molecular weight per electron and $m_u=1.66\times 10^{-24}
\text{g}$ the atomic mass unit. If the star is in a steady state, the
gravitational force balances the pressure force at every point. To
derive the {\sl hydrostatic equilibrium} equation, consider an infinitesimal
fluid element lying between $r$ and $r + dr$ and having an area $dA$   
perpendicular to the radial direction. The gravitational attraction
between $m(r)$ and the  mass $dm = \rho dAdr$ is the same as if $m(r)$
were concentrated in a point at the center, while the mass outside
exerts no force on $dm$. The net outward pressure force on $dm$ is
$-[P(r+dr)-P(r)]dA$, where $P$ is the pressure. So in equilibrium
\bq \label{es}
\frac{dP}{dr}=-\frac{Gm(r)\rho}{r^2}~,
\eq  
where $G$ is the universal gravitational constant. \footnote{Here we
  are assuming Newtonian theory of gravity. For the general
  relativistic stability analysis see for example \S 6.9 of
  Ref. \cite{Shapiro-Teukolsky}.}  

A consequence of the hydrostatic equilibrium is the {\sl virial
  theorem}. The gravitational potential energy of the star of radius
$R$ is
\bq \nonumber
W&=&-\int_0^R\frac{Gm(r)}{r}\rho 4\pi r^2\,dr\\ \nonumber
&=&\int_0^R\frac{dP}{dr}4\pi r^3\,dr\\ \label{virialt}
&=&-3\int_0^RP4\pi r^2\,dr~,
\eq  
where we have integrated by parts.

Now we assume that the gas of fermions is characterized by an
adiabatic equation of state
\bq \label{eos}
P=K\rho_0^\Gamma, ~~~K,\Gamma=1+\frac{1}{n}~\mbox{constants}~,
\eq
which is also called a {\sl polytrope} of polytropic index $n$.
For example for fermions in the extreme relativistic limit we find 
\bq \label{Kfer}
K=\frac{P}{\rho^{4/3}}=\frac{\pi^{2/3}\hbar c}{g^{1/3}(\mu_em_u)^{4/3}}
\frac{f_4(z)}{f_3^{4/3}(z)}~,
\eq
where $z$ depends on the temperature and density and goes to infinity
in the degenerate limit ($\lim_{z\to\infty}$ $f_4(z)/f_3^{4/3}(z)$ $=$
$3^{1/3}/2^{5/3}$). At the temperature and density typical of a white
dwarf $z$ is very large so the equation of state is practically
indistinguishable from the one in the degenerate limit.
  
Calling $u'$ the energy density of the gas, excluding the rest mass
energy, we must have from the first law of thermodynamics, assuming
adiabatic changes,
\bq
d(u/\rho_0)=-Pd(1/\rho_0)~,
\eq
and integration leads to
\bq
u=\rho_0c^2+\frac{P}{\Gamma-1}~,
\eq
which gives $u'=P/(\Gamma-1)$. Now Eq. (\ref{virialt}) can be
rewritten as 
\bq
W=-3(\Gamma-1)U~,
\eq
where $U=\int_0^Ru'4\pi r^2\,dr$ is the total internal energy of the
star. The total energy of the star, $E=W+U$, is then
\bq \label{tes}
E=-\frac{3\Gamma-4}{3(\Gamma-1)}|W|~.
\eq
If Eq. (\ref{eos}) holds everywhere inside the star of total mass $M$
and constant density, then the gravitational potential energy is given
by 
\bq \label{gpes}
W=-3\int_0^M\frac{P}{\rho}\,dm(r)=-\frac{3(\Gamma-1)}{5\Gamma}
\frac{GM^2}{R}~,
\eq
where we used $d(P/\rho)=[(\Gamma-1)/\Gamma]Gm(r)d(1/r)$ and
integrated by parts using $\Gamma>1$.

Without nuclear fuel, $E$ decreases due to radiation. According to
Eqs. (\ref{tes}) and (\ref{gpes}), $\Delta E<0$ implies $\Delta
R<0$ whenever $\Gamma>4/3$. That is the star contracts and the gas
will soon become quantum (see Ref. \cite{Shapiro-Teukolsky} \S 3.2).  
Can the star contract forever, extracting energy from the infinite
supply of gravitational potential energy until $R$ goes to zero or
until the star undergoes total collapse? The answer is no for stars
with $M \sim M_\odot$, as is demonstrated by Chandrasekhar
\cite{Chandrasekhar} or in the book of Shapiro and Teukolsky
\cite{Shapiro-Teukolsky}. We will reproduce their treatments in the
next section.  

\subsection{The Chandrasekhar limit}
\label{sec:Cl}
 
The hydrostatic equilibrium Eqs. (\ref{masss}) and (\ref{es}) can be
combined to give 
\bq
\frac{1}{r^2}\frac{d}{dr}\left(\frac{r^2}{\rho}
\frac{dP}{dr}\right)=-4\pi G\rho~.
\eq
Substituting the equation of state (\ref{eos}) and reducing the result
to dimensionless form with
\bq \label{nd1}
\rho&=&\rho_c\theta^n~,\\
r&=&a\eta~,\\
a&=&\sqrt{\frac{(n+1)K\rho_c^{1/n-1}}{4\pi G}}~,
\eq
where $\rho_c=\rho(r=0)$ is the central density, we find
\bq \label{le}
\frac{1}{\eta^2}\frac{d}{d\eta}\eta^2\frac{d\theta}{d\eta}=-\theta^n~.
\eq
This is the {\sl Lane-Emden equation} for the structure of a polytrope of
index $n$. The boundary conditions at the center of a polytropic star are 
\bq \label{lebc1}
\theta(0)&=&1~,\\ \label{lebc2}
\theta'(0)&=&0~.
\eq
The condition (\ref{lebc1}) follows directly from
Eq. (\ref{nd1}). Eq. (\ref{lebc2}) follows from the fact that near
the center $m(r)\approx4\pi\rho_cr^3/3$, so that by Eq. (\ref{masss})
$d\rho/dr=0$. 

Eq. (\ref{le}) can be easily integrated numerically, starting at
$\eta=0$ with the boundary conditions (\ref{lebc1}) and
(\ref{lebc2}). One finds that for $n<5$ ($\Gamma>6/5$), the solutions
decreases monotonically and have a zero at a finite value
$\eta=\eta_n$: $\theta(\eta_n)=0$. This point corresponds to the
surface of the star, where $P=\rho=0$. Thus the radius of the star is
\bq \label{le-radius}
R=a\eta_n~,
\eq
while the mass is
\bq \nonumber
M&=&\int_0^R4\pi r^2\rho\,dr\\ \nonumber 
&=&4\pi a^3\rho_c\int_0^{\eta_n}\eta^2\theta^n\,d\eta\\ \nonumber
&=&-4\pi a^3\rho_c\int_0^{\eta_n}\frac{d}{d\eta}
\left(\eta^2\frac{d\theta}{d\eta}\right)\,d\eta 
\\ \label{le-mass}
&=&4\pi a^3\rho_c\eta_n|\theta'(\eta_n)|~.
\eq
Eliminating $\rho_c$ between Eqs. (\ref{le-radius}) and
(\ref{le-mass}) gives the mass-radius relation for polytropes 
\bq
M=4\pi R^{(3-n)/(1-n)}\left[\frac{(n+1)K}{4\pi G}\right]^{n/(n-1)}
\eta_n^{(3-n)/(1-n)}\eta_n^2|\theta'(\eta_n)|~.
\eq
The solutions we are particularly interested in are
\bq
\Gamma&=&\frac{5}{3}~,~~~n=\frac{3}{2}~,~~~\eta_{3/2}=3.65375~,
~~~\eta_{3/2}^2|\theta'(\eta_{3/2})|=\omega_{3/2}=2.71406~,\\
\Gamma&=&\frac{4}{3}~,~~~n=3~,~~~\eta_3=6.89685~,
~~~\eta_3^2|\theta'(\eta_3)|=\omega_3=2.01824~,
\eq
which as explained in section \ref{sec:rehd} corresponds to the low
density non relativistic case and to the high density relativistic
case respectively.
Note that for $\Gamma=4/3$, $M$ is independent of $\rho_c$ and hence
$R$. We conclude that as $\rho_c\to\infty$, the electrons become more
and more relativistic throughout the star, and the mass asymptotically
approaches the value
\bq \label{Cl}
M_\text{Ch}=4\pi\omega_3\left(\frac{K}{\pi G}\right)^{3/2}~,
\eq
as $R\to 0$. The mass limit (\ref{Cl}) is called {\sl Chandrasekhar 
  limit} (see Eq. (36) in Re. \cite{Chandrasekhar1931}, Eq. (58) in
\cite{Chandrasekhar1935}, or Eq. (43) in \cite{Chandrasekhar1983}) and
represents the maximum possible mass of a white dwarf.

\begin{figure}[htbp]
\begin{center}
\includegraphics[width=10cm]{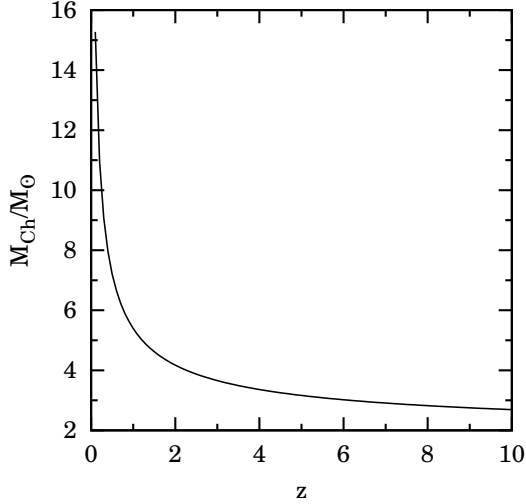}
\end{center}  
\caption{Temperature dependence of the Chandrasekar limit at
  $\mu_e=2$. We recall that $z=e^{\beta\mu}$.}  
\label{fig:mch-z}
\end{figure}

In Fig. \ref{fig:mch-z} we show the temperature dependence of the
Chandrasekar limit at $\mu_e=2$.

For the dependence of the star mass on the central density as it
develops through the various polytropes, as shown in
Fig. (\ref{fig:eos}), see for example Fig. 3.2 of
Ref. \cite{Shapiro-Teukolsky}. Clearly in the high 
$\rho_c\to\infty$ limit we will have in the degenerate limit
$z\to\infty$, from Eq. (\ref{Kfer}),
\bq
M\to M_\text{Ch}=1.45639\left(\frac{2}{\mu_e}\right)^2M_\odot~,
\eq
where $\mu_e$ can be taken approximately equal to $2$ or to $56/26$
assuming that all the elements have been subject to nuclear fusion
into the stable iron $\mbox{}^{56}_{26}$Fe.

The star will not become a black hole if $R>r_s$ (see Fig. 1.1 of
Ref. \cite{Shapiro-Teukolsky}), with $r_s=2GM_\text{Ch}/c^2$ the
Schwarzschild radius in the Chandrasekhar limit, i.e.
\bq \label{bh}
K<\frac{\eta_3c^2}{2^3\omega_3\rho_c^{1/3}}~,
\eq 
where $K$ is given by (\ref{Kfer}). This suggests that at high enough
central densities the star fate is to become a black hole. The
critical central density is given in the degenerate $z\to\infty$ limit
by $\bar{\rho}_c=g(\mu_e/2)^4(2.3542\times 10^{17}\text{g~cm}^{-3})$
which is well above the one required for the neutron drip.

If the star has a mass lower than $M_\text{Ch}$ it will not reach the
Chandrasekhar limit but will remain on a polytrope with $n<3$. If the
star has a mass higher than $M_\text{Ch}$ it will eventually evolve
through a supernovae explosion into a more compact object as a neutron
star (when electrons are captured by protons to form neutrons by
$\beta^+$ decay), a quark star, or a black hole. 

\section{The structure of the ideal quantum gas}
\label{sec:iqg-structure}

The radial distribution function $g(r)$ is related to the structure
factor $S(k)$ by the following Fourier transform
\bq \label{gr}
n[g(r)-1]=\frac{1}{V}\sum_{\kk}e^{i\kk\cdot\rr}[S(k)-1]~.
\eq
Taking into account that the operator of the particle number
$N_0$ is a constant of motion, the fluctuation-dissipation theorem 
(see appendix 5 of Ref. \cite{March-Tosi})
$\chi''(k,\omega)=(n\pi/\hbar)(1-e^{-\beta\hbar\omega})S(k,\omega)$, 
can be solved for the van Hove function
\bq
S(k,\omega)=\frac{\hbar}{n\pi}[1-\delta_\kk]
\frac{\chi''(k,\omega)}{1-e^{-\beta\hbar\omega}}+
\left\langle\frac{(\delta N)^2}{N}\right\rangle
\delta_\kk\delta(\omega)~,
\eq 
where $\langle\ldots\rangle$ represents averaging in the grand
canonical ensemble. The static structure factor
$S(k)=\int_{-\infty}^\infty d\omega\,S(k,\omega)$ then is
\bq
S(k)=\frac{\hbar}{n\pi}[1-\delta_\kk]
\int_0^\infty d\omega\,\chi''(k,\omega)\coth
\left(\frac{\beta\hbar\omega}{2}\right)+
\left\langle\frac{(\delta N)^2}{N}\right\rangle
\delta_\kk\delta(\omega)~,
\eq
where the last term does not contribute in the thermodynamic limit
\cite{Bosse2011}. We substitute (see appendix 8 of
Ref. \cite{March-Tosi}) 
\bq
\chi''(k,\omega)=N\pi\int\frac{d\kk'}{(2\pi)^3}\cc_{k'}
\{\delta[\hbar\omega-\Delta_{\kk'}(\kk)]-
\delta[\hbar\omega+\Delta_{\kk'}(\kk)]\}~,
\eq
with $\Delta_{\kk'}(\kk)=\epsilon(|\kk'+\kk|)-\epsilon(k')$, and
obtain for $\kk\neq\mathbf{0}$
\bq \label{sk}
S(k)=V\int\frac{d\kk'}{(2\pi)^3}\cc_{k'}\coth\left\{\frac{1}{2}
\beta[\epsilon(|\kk'+\kk|)-\epsilon(k')]\right\}~,~~~k>0~,
\eq
where $\cc_k$ denotes the thermal average fraction of particles having
momentum $\hbar\kk$ defined in Eq. (\ref{ck}).

\begin{figure}[htbp]
\begin{center}
\includegraphics[width=10cm]{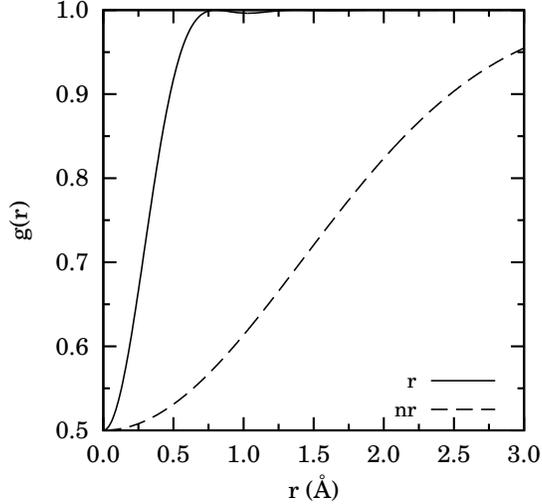}
\end{center}  
\caption{The radial distribution function for ideal electrons
  ($\xi=-1, g=2$) in the relativistic and the non relativistic
  cases. Here we chose $T=20000\text{K}$ and $n=1.04\times 10^{22}
  \text{cm}^{-3}$ in the non relativistic case and $n=5.93\times 10^{24}
  \text{cm}^{-3}$ in the relativistic case. $r$ is in Angstroms.}
\label{fig:gr}
\end{figure}

For further analytical manipulation we rewrite
\bq \label{ck1}
\frac{\beta}{2}[\epsilon(k)-\mu]=\ln\sqrt{\frac{g}{N\cc_k}+\xi}~,
\eq
One rewrites Eq. (\ref{sk}) changing variables first $\kk+\kk'\to\kk$
and subsequently $\kk\to-\kk$ to find
\bq \label{sk1}
S(k)=V\int\frac{d\kk'}{(2\pi)^3}\cc_{|\kk+\kk'|}\coth\left\{\frac{1}{2}
\beta[\epsilon(k)-\epsilon(|\kk+\kk'|)]\right\}~.
\eq
Adding Eqs. (\ref{sk}) and (\ref{sk1}) and making use of the fact
that the hyperbolic cotangent is an odd function, one finds
\bq
2S(k)=V\int\frac{d\kk'}{(2\pi)^3}(\cc_{k'}-\cc_{|\kk+\kk'|})
\coth\left\{\frac{1}{2}
\beta[\epsilon(|\kk'+\kk|)-\epsilon(k')]\right\}~.
\eq
Now using Eq. (\ref{ck1}) we find
\bq \nonumber
S(k)&=&\frac{V}{2}\int\frac{d\kk'}{(2\pi)^3}\,(\cc_{k'}-\cc_{|\kk+\kk'|})
\coth\left[\ln\sqrt{\frac{g}{N\cc_{|\kk+\kk'|}}+\xi}-
\ln\sqrt{\frac{g}{N\cc_{k'}}+\xi}\right]\\ \nonumber
&=&\frac{V}{2}\int\frac{d\kk'}{(2\pi)^3}
\left(\cc_{k'}+\cc_{|\kk+\kk'|}+\frac{2N\xi}{g}\cc_{k'}\cc_{|\kk+\kk'|}\right)\\
\label{sk2}
&=&1+\frac{VN\xi}{g}\int\frac{d\kk'}{(2\pi)^3}
\cc_{k'}\cc_{|\kk+\kk'|}~,~~~k>0~,
\eq
where $\coth[\ln\sqrt{x}]=(x+1)/(x-1)$ was used in the middle
step. From this follows
\bq
\frac{1}{V}\sum_{\kk\neq\mathbf{0}}e^{i\kk\cdot\rr}[S(k)-1]=
\frac{n\xi}{g}\left\{2\cc_0\sum_{\kk\neq\mathbf{0}}\,\cc_ke^{i\kk\cdot\rr}
+\left|\sum_{\kk\neq\mathbf{0}}\,\cc_ke^{i\kk\cdot\rr}\right|^2\right\}~,
\eq
where $\cc_0=\delta_{\xi,1}\Theta(T_c-T)N_0/N$, with $\Theta$ the
Heaviside step function, denotes the fraction of particles which
occupy the zero momentum state. We then introduce the function 
$F(r)=\sum_{\kk}\cc_{k}e^{i\kk\cdot\rr}$. This assume the following
forms  
\bq
F_{r}(r)&=&
\cc_0(T)+\frac{g}{2\pi^2n(\beta\hbar c)^2\xi}\int_0^\infty
\kappa d\kappa\, b_0\left(\xi z e^{-\sqrt{\kappa^2+\beta^2 m^2c^4}}\right)
\sin\left(\frac{1}{\beta\hbar c}\kappa r\right)/r~,\\
F_{er}(r)&=&
\cc_0(T)+\frac{g}{2\pi^2n(\beta\hbar c)^2\xi}\int_0^\infty
\kappa d\kappa\, b_0\left(\xi z e^{-\kappa}\right)
\sin\left(\frac{1}{\beta\hbar c}\kappa r\right)/r~,\\
F_{nr}(r)&=&
\cc_0(T)+\frac{2g}{\pi n\Lambda^2\xi}\int_0^\infty
\kappa d\kappa\, b_0\left(\xi z e^{-\kappa^2}\right)
\sin\left(\frac{2\sqrt{\pi}}{\Lambda}\kappa r\right)/r~.
\eq
in the relativistic $\epsilon(k)=\sqrt{\hbar^2k^2c^2+m^2c^4}$, extreme
relativistic $\epsilon(k)=c\hbar k$, and non relativistic
$\epsilon(k)=\lambda k^2$ cases respectively. Inserting
Eq. (\ref{sk2}) into Eq. (\ref{gr}) we find 
\bq
g(r)=1+\frac{\xi}{g}\left[F^2(r)-\cc_0^2(T)\right]~.
\eq
which generalizes Eq. (117.8) of Landau \cite{LandauCTP5}. In
Fig. \ref{fig:gr} we show the redial distribution function for
fermions in the relativistic and the non relativistic cases. From the
figure we see how the Fermi hole becomes larger in the non
relativistic case at smaller number densities. Increasing the
temperature by one order of magnitude (see Fig. 3.3 of
Ref. \cite{Shapiro-Teukolsky}) keeping the density fixed 
produces a change in the redial distribution function of the order of
$10^{-2}$, with the Fermi hole getting smaller. 

For the electron gas we should include the Coulomb interaction between
the particles: the {\sl jellium}. The radial distribution function of
jellium cannot of course be calculated exactly analytically, for a
Monte Carlo simulation of the degenerate ($T=0$) jellium see for
example Ref. \cite{Fantoni13g} and for jellium at finite temperature
see for example Ref. \cite{Militzer2003}. 

Actually a more accurate result could be found by treating the white
dwarf matter as a binary mixture of electrons and nuclei which can
today be done exactly with Monte Carlo simulations techniques like the
one devised in Ref. \cite{Dewing2002}. 

From these numerical studies one could extract a more accurate value
for the constant $K$ in the adiabatic equation of state and thus 
the critical central density $\bar{\rho}_c=(\eta_3 c^2/2^3\omega_3K)^3$. 

\section{Conclusions} 
\label{app:conclusions} 

In this work we studied the importance of temperature dependence on
ideal Quantum gases relevant for white dwarfs interior. Even if the
temperature of the star is six orders of magnitudes smaller
than the Fermi energy of the electron gas inside the star, we find
that the temperature effects are quite relevant at the white dwarf
densities and temperatures. In particular we show that the adiabatic
equation of state 
becomes extremely relativistic, with $\Gamma=4/3$, at densities six
orders of magnitude lower than the ones required for the completely
degenerate, $T=0$, case. Even if the polytropic form of the
adiabatic equation of state remains the same as that at zero
temperature, the proportionality constant $K$ changing by just a
$10^{-10}$ relative factor between the finite temperature case and the
zero temperature case, we think that an accurate analysis of the
star evolution, at least at the level of the ideal electron gas
approximation in absence of the nuclei, should properly take into
account the temperature effects. This gives us a complete exactly
solvable analytic approximation for the compact star interior at a
finite temperature. We could comment that the temperature effects are
smaller than the corrections necessary to take into account of the
Coulomb interactions between the electrons and of the presence of the
nuclei, but from a calculation point of view it is still desirable to
keep under control the magnitude of the temperature corrections
alone. Since this can be done analytically we think that their
analysis is relevant by itself.

We gave the generalization to finite temperature of all the zero
temperature results used by Chandrasekhar and in order to keep the
treatment as general as possible we studied in parallel the Fermi and
the Bose gas. Clearly only the Fermi gas results were used for the
description of the ideal electron gas in the star interior.

We then studied the structure of the ideal quantum gas as a function
of temperature. We found the Fermi hole for the cold electron gas in a
white dwarf which turned out to be of the order of 1\r{A} in the
full relativistic regime at a number density of the order of $n\sim
10^{26}\text{cm}^{-3}$ and bigger in the non relativistic regime at
smaller densities and fixed temperature. The radial distribution
function is also affected by the temperature and the Fermi hole gets
smaller as the temperature increases at fixed density.

We also point out that in order to correct our result for the Coulomb
interaction among the electrons and for the presence of the nuclei, it
is necessary to abandon the analytic treatment in favor of the
numerical simulation. We gave some relevant references of Monte Carlo
methods which are important to adopt to solve this fascinating
subject. These corrections to the Chandrasekhar result or to our
temperature dependent treatment are important more from a
philosophical point of view rather than an experimental or
observational point of view. They would lead us to the exact knowledge
of the properties of a mixture of electrons and nuclei at
astrophysical conditions such as the ones found in white dwarfs.   

More over let us observe that only a general relativistic statistical
physics theory would give us fully correct results for the stability
of a white dwarf. But since this theory has not yet been formulated
\cite{Rovelli2013} we will have to wait till the theory becomes
available. 

\appendix
\section{The adiabatic equation of state for a relativistic ideal
  electron gas at finite temperature}  
\label{app:1} 

Using the dispersion relation
$\epsilon(k)=\sqrt{\hbar^2k^2c^2+m^2c^4}$, with $m$ the rest mass of
an electron, we find the pressure and the density from,
\bq \label{P}
\beta P&=&g\int\frac{d\kk}{(2\pi)^3}\ln\left(
1+ze^{-\beta\epsilon(k)}\right)~,\\
n&=&g\int\frac{d\kk}{(2\pi)^3}\frac{1}{e^{\beta\epsilon(k)}/z+1}~.
\eq
Integrating by parts the pressure equation and changing variable
$\kappa=\beta\hbar c k$ we find
\bq
\beta P&=&\frac{g}{(\beta\hbar c)^3}\frac{1}{2\pi^2}\frac{1}{3}
\int d\kappa\,\frac{\kappa^3/\sqrt{\kappa^2+(\beta mc^2)^2}}
{e^{\sqrt{\kappa^2+(\beta mc^2)^2}}/z+1}~,\\
n&=&\frac{g}{(\beta\hbar c)^3}\frac{1}{2\pi^2}
\int d\kappa\,\frac{\kappa^2}{e^{\sqrt{\kappa^2+(\beta mc^2)^2}}/z+1}~.
\eq
These equations are equivalent to Eqs. (\ref{Prf}) and (\ref{denrf})
in the main text. Then the entropy is given by
\bq
S/Vk_B=g\int\frac{d\kk}{(2\pi)^3}\ln\left(
1+ze^{-\beta\epsilon(k)}\right)-g\int\frac{d\kk}{(2\pi)^3}
\frac{\ln z-\beta\epsilon(k)}{e^{\beta\epsilon(k)}/z+1}~.
\eq
On an adiabatic the entropy per particle $s=S/Nk_B$ is constant, and from
Eq. (\ref{P}) follows
\bq
\beta P=
g\int\frac{d\kk}{(2\pi)^3}
\frac{\ln z-\beta\epsilon(k)}{e^{\beta\epsilon(k)}/z+1}+sn~.
\eq
 

\begin{thebibliography}{25}%
\makeatletter
\providecommand \@ifxundefined [1]{%
 \@ifx{#1\undefined}
}%
\providecommand \@ifnum [1]{%
 \ifnum #1\expandafter \@firstoftwo
 \else \expandafter \@secondoftwo
 \fi
}%
\providecommand \@ifx [1]{%
 \ifx #1\expandafter \@firstoftwo
 \else \expandafter \@secondoftwo
 \fi
}%
\providecommand \natexlab [1]{#1}%
\providecommand \enquote  [1]{``#1''}%
\providecommand \bibnamefont  [1]{#1}%
\providecommand \bibfnamefont [1]{#1}%
\providecommand \citenamefont [1]{#1}%
\providecommand \href@noop [0]{\@secondoftwo}%
\providecommand \href [0]{\begingroup \@sanitize@url \@href}%
\providecommand \@href[1]{\@@startlink{#1}\@@href}%
\providecommand \@@href[1]{\endgroup#1\@@endlink}%
\providecommand \@sanitize@url [0]{\catcode `\\12\catcode `\$12\catcode
  `\&12\catcode `\#12\catcode `\^12\catcode `\_12\catcode `\%12\relax}%
\providecommand \@@startlink[1]{}%
\providecommand \@@endlink[0]{}%
\providecommand \url  [0]{\begingroup\@sanitize@url \@url }%
\providecommand \@url [1]{\endgroup\@href {#1}{\urlprefix }}%
\providecommand \urlprefix  [0]{URL }%
\providecommand \Eprint [0]{\href }%
\providecommand \doibase [0]{http://dx.doi.org/}%
\providecommand \selectlanguage [0]{\@gobble}%
\providecommand \bibinfo  [0]{\@secondoftwo}%
\providecommand \bibfield  [0]{\@secondoftwo}%
\providecommand \translation [1]{[#1]}%
\providecommand \BibitemOpen [0]{}%
\providecommand \bibitemStop [0]{}%
\providecommand \bibitemNoStop [0]{.\EOS\space}%
\providecommand \EOS [0]{\spacefactor3000\relax}%
\providecommand \BibitemShut  [1]{\csname bibitem#1\endcsname}%
\let\auto@bib@innerbib\@empty
\bibitem [{\citenamefont {Balian}\ and\ \citenamefont
  {Blaizot}(1999)}]{Balian99}%
  \BibitemOpen
  \bibfield  {author} {\bibinfo {author} {\bibfnamefont {R.}~\bibnamefont
  {Balian}}\ and\ \bibinfo {author} {\bibfnamefont {J.~P.}\ \bibnamefont
  {Blaizot}},\ }\href@noop {} {\bibfield  {journal} {\bibinfo  {journal} {Am.
  J. Phys.}\ }\textbf {\bibinfo {volume} {67}},\ \bibinfo {pages} {1189}
  (\bibinfo {year} {1999})}\BibitemShut {NoStop}%
\bibitem [{\citenamefont {Silbar}\ and\ \citenamefont
  {Reddy}(2004)}]{Silbar04}%
  \BibitemOpen
  \bibfield  {author} {\bibinfo {author} {\bibfnamefont {R.}~\bibnamefont
  {Silbar}}\ and\ \bibinfo {author} {\bibfnamefont {S.}~\bibnamefont {Reddy}},\
  }\href@noop {} {\bibfield  {journal} {\bibinfo  {journal} {Am. J. Phys.}\
  }\textbf {\bibinfo {volume} {72}},\ \bibinfo {pages} {892} (\bibinfo {year}
  {2004})}\BibitemShut {NoStop}%
\bibitem [{\citenamefont {Jackson}\ \emph {et~al.}(2005)\citenamefont
  {Jackson}, \citenamefont {Taruna}, \citenamefont {Pouliot}, \citenamefont
  {Ellison}, \citenamefont {Lee},\ and\ \citenamefont
  {Piekarewicz}}]{Jackson05}%
  \BibitemOpen
  \bibfield  {author} {\bibinfo {author} {\bibfnamefont {C.~B.}\ \bibnamefont
  {Jackson}}, \bibinfo {author} {\bibfnamefont {J.}~\bibnamefont {Taruna}},
  \bibinfo {author} {\bibfnamefont {S.~L.}\ \bibnamefont {Pouliot}}, \bibinfo
  {author} {\bibfnamefont {B.~W.}\ \bibnamefont {Ellison}}, \bibinfo {author}
  {\bibfnamefont {D.~D.}\ \bibnamefont {Lee}}, \ and\ \bibinfo {author}
  {\bibfnamefont {J.}~\bibnamefont {Piekarewicz}},\ }\href@noop {} {\bibfield
  {journal} {\bibinfo  {journal} {European J. Phys.}\ }\textbf {\bibinfo
  {volume} {26}},\ \bibinfo {pages} {695} (\bibinfo {year} {2005})}\BibitemShut
  {NoStop}%
\bibitem [{\citenamefont {Shapiro}\ and\ \citenamefont
  {Teukolsky}(1983)}]{Shapiro-Teukolsky}%
  \BibitemOpen
  \bibfield  {author} {\bibinfo {author} {\bibfnamefont {S.~L.}\ \bibnamefont
  {Shapiro}}\ and\ \bibinfo {author} {\bibfnamefont {S.~A.}\ \bibnamefont
  {Teukolsky}},\ }\href@noop {} {\emph {\bibinfo {title} {Black Holes, White
  Dwarfs, and Neutro Stars, the physics of compact objects}}}\ (\bibinfo
  {publisher} {John Wiley \& Sons, Inc.},\ \bibinfo {address} {New York},\
  \bibinfo {year} {1983})\BibitemShut {NoStop}%
\bibitem [{\citenamefont {Landau}(1932)}]{Landau1932}%
  \BibitemOpen
  \bibfield  {author} {\bibinfo {author} {\bibfnamefont {L.~D.}\ \bibnamefont
  {Landau}},\ }\href@noop {} {\bibfield  {journal} {\bibinfo  {journal} {Phys.
  Z. Sowjetunion}\ }\textbf {\bibinfo {volume} {1}},\ \bibinfo {pages} {285}
  (\bibinfo {year} {1932})}\BibitemShut {NoStop}%
\bibitem [{\citenamefont
  {Chandrasekhar}(1931{\natexlab{a}})}]{Chandrasekhar1931}%
  \BibitemOpen
  \bibfield  {author} {\bibinfo {author} {\bibfnamefont {S.}~\bibnamefont
  {Chandrasekhar}},\ }\href@noop {} {\bibfield  {journal} {\bibinfo  {journal}
  {Monthly Notices of the Royal Astronomical Society}\ }\textbf {\bibinfo
  {volume} {91}},\ \bibinfo {pages} {456} (\bibinfo {year}
  {1931}{\natexlab{a}})}\BibitemShut {NoStop}%
\bibitem [{\citenamefont
  {Chandrasekhar}(1931{\natexlab{b}})}]{Chandrasekhar1931a}%
  \BibitemOpen
  \bibfield  {author} {\bibinfo {author} {\bibfnamefont {S.}~\bibnamefont
  {Chandrasekhar}},\ }\href@noop {} {\bibfield  {journal} {\bibinfo  {journal}
  {Phil. Mag.}\ }\textbf {\bibinfo {volume} {11}},\ \bibinfo {pages} {592}
  (\bibinfo {year} {1931}{\natexlab{b}})}\BibitemShut {NoStop}%
\bibitem [{\citenamefont
  {Chandrasekhar}(1931{\natexlab{c}})}]{Chandrasekhar1931b}%
  \BibitemOpen
  \bibfield  {author} {\bibinfo {author} {\bibfnamefont {S.}~\bibnamefont
  {Chandrasekhar}},\ }\href@noop {} {\bibfield  {journal} {\bibinfo  {journal}
  {Astrophys. J.}\ }\textbf {\bibinfo {volume} {74}},\ \bibinfo {pages} {81}
  (\bibinfo {year} {1931}{\natexlab{c}})}\BibitemShut {NoStop}%
\bibitem [{\citenamefont {Dirac}(1926)}]{Dirac1926}%
  \BibitemOpen
  \bibfield  {author} {\bibinfo {author} {\bibfnamefont {P.~A.~M.}\
  \bibnamefont {Dirac}},\ }\href@noop {} {\bibfield  {journal} {\bibinfo
  {journal} {Proc. Roy. Soc. London. Ser. A}\ }\textbf {\bibinfo {volume}
  {112}},\ \bibinfo {pages} {661} (\bibinfo {year} {1926})}\BibitemShut
  {NoStop}%
\bibitem [{\citenamefont {Fowler}(1926)}]{Fowler1926}%
  \BibitemOpen
  \bibfield  {author} {\bibinfo {author} {\bibfnamefont {R.~H.}\ \bibnamefont
  {Fowler}},\ }\href@noop {} {\bibfield  {journal} {\bibinfo  {journal} {Mon.
  Not. Roy. Astron. Soc.}\ }\textbf {\bibinfo {volume} {87}},\ \bibinfo {pages}
  {114} (\bibinfo {year} {1926})}\BibitemShut {NoStop}%
\bibitem [{\citenamefont {Adams}(1915)}]{Adams1915}%
  \BibitemOpen
  \bibfield  {author} {\bibinfo {author} {\bibfnamefont {W.~S.}\ \bibnamefont
  {Adams}},\ }\href@noop {} {\bibfield  {journal} {\bibinfo  {journal} {Pub.
  Astron. Soc. Pac.}\ }\textbf {\bibinfo {volume} {27}},\ \bibinfo {pages}
  {236} (\bibinfo {year} {1915})}\BibitemShut {NoStop}%
\bibitem [{\citenamefont {Adams}(1925)}]{Adams1925}%
  \BibitemOpen
  \bibfield  {author} {\bibinfo {author} {\bibfnamefont {W.~S.}\ \bibnamefont
  {Adams}},\ }\href@noop {} {\bibfield  {journal} {\bibinfo  {journal} {Proc.
  Natl. Acad. Sci. USA}\ }\textbf {\bibinfo {volume} {11}},\ \bibinfo {pages}
  {382} (\bibinfo {year} {1925})},\ \bibinfo {note} {erratum: {\sl Observatory}
  {\bf 49}, 88}\BibitemShut {NoStop}%
\bibitem [{Note1()}]{Note1}%
  \BibitemOpen
  \bibinfo {note} {This value will get smaller as the star cools down in view
  of Eq. (\ref {denerf}) and eventually become close to zero as the momentum
  theraml average fraction approaches a step function}\BibitemShut {NoStop}%
\bibitem [{\citenamefont {Feynman}(1972)}]{FeynmanFIP}%
  \BibitemOpen
  \bibfield  {author} {\bibinfo {author} {\bibfnamefont {R.~P.}\ \bibnamefont
  {Feynman}},\ }\href@noop {} {\emph {\bibinfo {title} {Statistical Mechanics:
  A Set of Lectures}}},\ \bibinfo {series} {Frontiers in Physics},
  Vol.~\bibinfo {volume} {36}\ (\bibinfo  {publisher} {W. A. Benjamin, Inc.},\
  \bibinfo {year} {1972})\ \bibinfo {note} {notes taken by R. Kikuchi and H. A.
  Feiveson, edited by Jacob Shaham}\BibitemShut {NoStop}%
\bibitem [{\citenamefont {Landau}\ and\ \citenamefont
  {Lifshitz}(1980)}]{LandauCTP5}%
  \BibitemOpen
  \bibfield  {author} {\bibinfo {author} {\bibfnamefont {L.~D.}\ \bibnamefont
  {Landau}}\ and\ \bibinfo {author} {\bibfnamefont {E.~M.}\ \bibnamefont
  {Lifshitz}},\ }\href@noop {} {\emph {\bibinfo {title} {Statistical Physics,
  Part I}}},\ \bibinfo {edition} {3rd}\ ed.,\ \bibinfo {series} {Course of
  Theoretical Physics}, Vol.~\bibinfo {volume} {5}\ (\bibinfo  {publisher}
  {Butterworth Heinemann},\ \bibinfo {year} {1980})\ \bibinfo {note}
  {translated from the Russian by J. B. Sykes and M. J. Kearsley, edited by E.
  M. Lifshitz and L. P. Pitaevskii}\BibitemShut {NoStop}%
\bibitem [{Note2()}]{Note2}%
  \BibitemOpen
  \bibinfo {note} {Here we are assuming Newtonian theory of gravity. For the
  general relativistic stability analysis see for example \protect \S 6.9 of
  Ref. \cite {Shapiro-Teukolsky}.}\BibitemShut {Stop}%
\bibitem [{\citenamefont {Chandrasekhar}(1938)}]{Chandrasekhar}%
  \BibitemOpen
  \bibfield  {author} {\bibinfo {author} {\bibfnamefont {S.}~\bibnamefont
  {Chandrasekhar}},\ }\href@noop {} {\emph {\bibinfo {title} {An introduction
  to the study of stellar structure}}}\ (\bibinfo  {publisher} {The University
  of Chicago Press},\ \bibinfo {address} {Chicago, Illinois},\ \bibinfo {year}
  {1938})\BibitemShut {NoStop}%
\bibitem [{\citenamefont {Chandrasekhar}(1935)}]{Chandrasekhar1935}%
  \BibitemOpen
  \bibfield  {author} {\bibinfo {author} {\bibfnamefont {S.}~\bibnamefont
  {Chandrasekhar}},\ }\href@noop {} {\bibfield  {journal} {\bibinfo  {journal}
  {Monthly Notices of the Royal Astronomical Society}\ }\textbf {\bibinfo
  {volume} {95}},\ \bibinfo {pages} {207} (\bibinfo {year} {1935})}\BibitemShut
  {NoStop}%
\bibitem [{\citenamefont {Chandrasekhar}(1983)}]{Chandrasekhar1983}%
  \BibitemOpen
  \bibfield  {author} {\bibinfo {author} {\bibfnamefont {S.}~\bibnamefont
  {Chandrasekhar}},\ }\href@noop {} {\bibfield  {journal} {\bibinfo  {journal}
  {Nobel Prize lecture}\ } (\bibinfo {year} {1983})}\BibitemShut {NoStop}%
\bibitem [{\citenamefont {March}\ and\ \citenamefont
  {Tosi}(1984)}]{March-Tosi}%
  \BibitemOpen
  \bibfield  {author} {\bibinfo {author} {\bibfnamefont {N.~H.}\ \bibnamefont
  {March}}\ and\ \bibinfo {author} {\bibfnamefont {M.~P.}\ \bibnamefont
  {Tosi}},\ }\href@noop {} {\emph {\bibinfo {title} {Coulomb liquids}}}\
  (\bibinfo  {publisher} {Academic Press},\ \bibinfo {year} {1984})\BibitemShut
  {NoStop}%
\bibitem [{\citenamefont {Bosse}\ \emph {et~al.}(2011)\citenamefont {Bosse},
  \citenamefont {Pathak},\ and\ \citenamefont {Singh}}]{Bosse2011}%
  \BibitemOpen
  \bibfield  {author} {\bibinfo {author} {\bibfnamefont {J.}~\bibnamefont
  {Bosse}}, \bibinfo {author} {\bibfnamefont {K.~N.}\ \bibnamefont {Pathak}}, \
  and\ \bibinfo {author} {\bibfnamefont {G.~S.}\ \bibnamefont {Singh}},\
  }\href@noop {} {\bibfield  {journal} {\bibinfo  {journal} {Phys. Rev. E}\
  }\textbf {\bibinfo {volume} {84}},\ \bibinfo {pages} {042101} (\bibinfo
  {year} {2011})}\BibitemShut {NoStop}%
\bibitem [{\citenamefont {Fantoni}(2013)}]{Fantoni13g}%
  \BibitemOpen
  \bibfield  {author} {\bibinfo {author} {\bibfnamefont {R.}~\bibnamefont
  {Fantoni}},\ }\href@noop {} {\bibfield  {journal} {\bibinfo  {journal} {Eur.
  Phys. J. B}\ }\textbf {\bibinfo {volume} {86}},\ \bibinfo {pages} {286}
  (\bibinfo {year} {2013})}\BibitemShut {NoStop}%
\bibitem [{\citenamefont {Militzer}\ \emph {et~al.}(2003)\citenamefont
  {Militzer}, \citenamefont {Pollock},\ and\ \citenamefont
  {Ceperley}}]{Militzer2003}%
  \BibitemOpen
  \bibfield  {author} {\bibinfo {author} {\bibfnamefont {B.}~\bibnamefont
  {Militzer}}, \bibinfo {author} {\bibfnamefont {E.~L.}\ \bibnamefont
  {Pollock}}, \ and\ \bibinfo {author} {\bibfnamefont {D.~M.}\ \bibnamefont
  {Ceperley}},\ }\href@noop {} {\  (\bibinfo {year} {2003})},\ \bibinfo {note}
  {cond-mat/0310401}\BibitemShut {NoStop}%
\bibitem [{\citenamefont {Dewing}\ and\ \citenamefont
  {Ceperley}(2002)}]{Dewing2002}%
  \BibitemOpen
  \bibfield  {author} {\bibinfo {author} {\bibfnamefont {M.}~\bibnamefont
  {Dewing}}\ and\ \bibinfo {author} {\bibfnamefont {D.~M.}\ \bibnamefont
  {Ceperley}},\ }in\ \href@noop {} {\emph {\bibinfo {booktitle} {Recent
  Advances in Quantum Monte Carlo Methods, II}}},\ \bibinfo {editor} {edited
  by\ \bibinfo {editor} {\bibfnamefont {W.~A.}\ \bibnamefont {Lester}},
  \bibinfo {editor} {\bibfnamefont {S.~M.}\ \bibnamefont {Rothstein}}, \ and\
  \bibinfo {editor} {\bibfnamefont {S.}~\bibnamefont {Tanaka}}}\ (\bibinfo
  {publisher} {World Scientific},\ \bibinfo {address} {Singapore},\ \bibinfo
  {year} {2002})\BibitemShut {NoStop}%
\bibitem [{\citenamefont {Rovelli}(2013)}]{Rovelli2013}%
  \BibitemOpen
  \bibfield  {author} {\bibinfo {author} {\bibfnamefont {C.}~\bibnamefont
  {Rovelli}},\ }\href@noop {} {\bibfield  {journal} {\bibinfo  {journal} {Phys.
  Rev. D}\ }\textbf {\bibinfo {volume} {87}},\ \bibinfo {pages} {084055}
  (\bibinfo {year} {2013})}\BibitemShut {NoStop}%
\end{thebibliography}%

%

\end{document}